# Modeling and Analysis of SLED

LI Lin[1,2] , FANG WenCheng[1] ,WANG Chao-Peng[1] ,GU Qiang[1*]

[1] *Shanghai Institute of Applied Physics, Chinese Academy of Sciences, Shanghai 201800, China*

[2] *University of Chinese Academy of Science, Beijing 100049, China*

**Abstract** SLED is a crucial component for C-band microwave acceleration unit of SXFEL. To study the behavior of SLED (SLAC Energy Doubler), mathematic model is commonly built and analyzed. In this paper, a new method is proposed to build the model of SLED at SINAP. With this method, the parameters of the two cavities can be analyzed separately. Also it is suitable to study parameter optimization of SLED and analyze the effect from the parameters variations. Simulation results of our method are also presented.

* Corresponding author(email: guqiang@sinap.ac.cn)

## 1 Introduction

A compact soft X-ray Free Electron Laser (SXFEL) facility is presently planned at Shanghai Institute of Applied Physics, CAS [1], and some analytical modeling and simulation research is ongoing. The high power RF system for SXFEL comprises a RF power source, a constant gradient accelerating structure and waveguide components For getting a high constant gradient field in the accelerating structure, the existing klystron power source of 50 MW cannot meet the power requirement of the field target, and a pulse compressor is required to multiply the power from the klystron [2]. There are different types of pulse compressor which satisfies the requirements. In our case, a SLED type pulse compressor is proposed for the C-band RF system in SXFEL.

To study the performance of the pulse compressor and analyze the parameters, an effective way is building a mathematic model, then the model based simulations can be implemented to verify the design. In this paper, a mathematic model of the SLED presented, which is a powerful tool for control system development. With this model, the parameters of SLED are optimized and the effects of the parameters variations are analyzed correspondingly.

## 2 Modeling of the SLED

### 2.1 Structure of the SLED

SLED is a RF pulse compressor which was firstly invented by Farkers Z at 1974[4].The SLED is composed of two identical high Q-factor cavities attached to a 3 dB coupler. The structure of SLED is shown in Fig. 1.

The performance of SLED is determined by the structure of the storage cavities. The energy multiplication factor $M$ can be expressed as:

$$M = \gamma e^{-\frac{T_a}{T_c}}\left[1-(1+g)^{1+v}\right]\left[g(1+v)\right]^{-1} - (\alpha - 1) \quad (1)$$

Where $T_a$ is the filling time of the accelerating structure, $T_c$ is the filling time of the cavity. $\alpha = 2\beta/(\beta+1)$, $v = T_a/(T_c \ln(1-g)) - 1$, $g$ is the gradient of the group velocity along the accelerating structure. $\gamma = \beta(2 - e^{-T_1})$, $\beta$ is the cavity coupling coefficient.

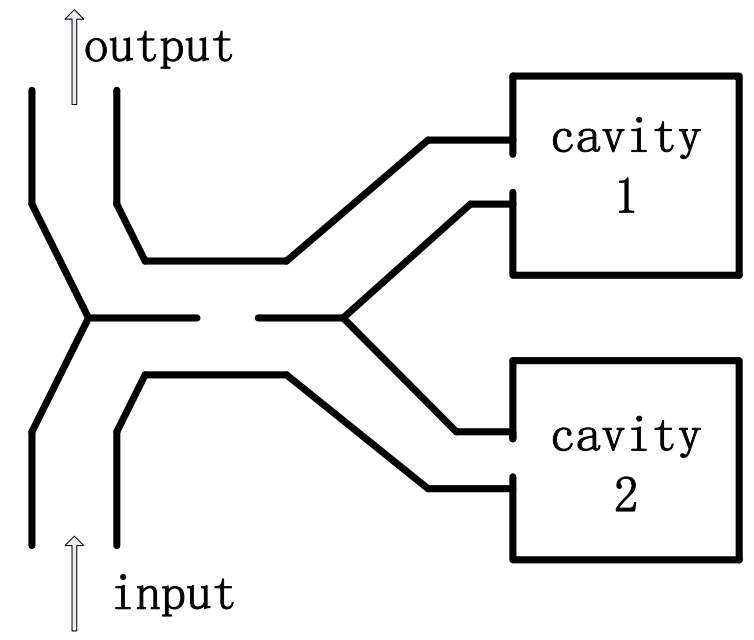

**Fig. 1** The structure of the SLED

### 2.2 Modeling SLED using S11

Based on references [3][4][5], a model of SLED can be constructed by the energy conservation. However the models based on these methods contain only the amplitude information of the input and output signal, no phase information can be reflected, and not suitable for the case of two asymmetry cavities.

In the paper the technology of two ports terminal network is used to model the behavior of the cavity and input coupler. The cavity itself is equivalent to a RLC circuit, and the input coupler can be presented by an ideal transformer as show in Fig. 2.

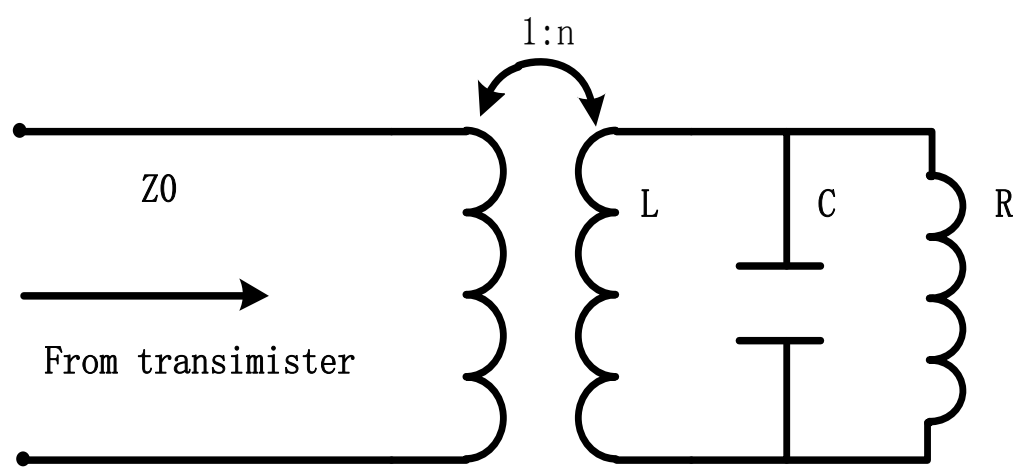

**Fig. 2** The equivalent circuit of SLED

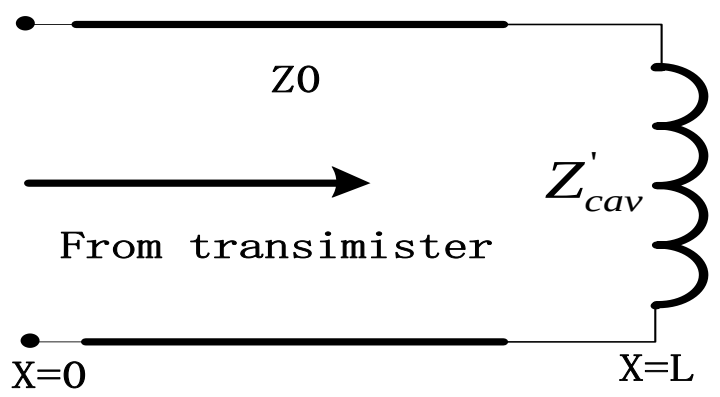

**Fig. 3** Cvity model at the transmission line side.

Using the definition of reflection coefficient for two ports microwave network in Fig. 3, the formula of reflection coefficient can be presented as [6]:

$$S_{11} = \frac{V_{out}}{V_{in}} = \frac{Z'_{cav} - Z_0}{Z'_{cav} + Z_0} \qquad (2)$$

Where $Z'_{cav} = \beta Z_0 \frac{2\sigma s}{s^2 + 2\sigma s + \omega_0^2}$ , $\sigma = \frac{\omega_0}{2Q_0}$ , $Q_0 = \omega_0 RC$ is the cavity unloaded quality factor and $\beta = \frac{Q_0}{Q_{ext}} = \frac{R}{n^2 Z_0}$ is the coupling coefficient.

Substituting the impedance $Z'_{cav}$ into equation (2), then equation (2) can be represented by the differential equation between the input and output voltage

$$\frac{d^2V_{out}}{dt^2} + 2\sigma(\beta+1)\frac{dV_{out}}{dt} + \omega_0^2 V_{out} = -\frac{d^2V_{in}}{dt^2} + 2\sigma(\beta-1)\frac{dV_{in}}{dt} - \omega_0^2 V_{in} \qquad (3)$$

The input voltage $V_{in}$ and output voltage $V_{out}$ are modulated sine wave with the frequency of $\omega$, the phase and amplitude information is the most useful part that we should consider, so, they can be written into phasor as

$$V_{in} = \overrightarrow{V_{in}} e^{j\omega t}, V_{out} = \overrightarrow{V_{out}} e^{i\omega t}$$

Where the $\overrightarrow{V_{in}}$ and $\overrightarrow{V_{out}}$ are the complex amplitude as vectors which contain the amplitude and phase information, which can be expressed by their real and image parts. Put the phasor definition into the equation above and assume $\sigma(\beta-1) << \omega$, and define the detuning as $\Delta\omega = \omega_0 - \omega$ . If the detuning is small than working frequency, we will get the approximation of $\omega_0^2 - \omega^2 = 2\omega\Delta\omega$ . Normally, the voltage changes slowly, so the item for the second derivation is always smaller than others. So the equation (3) can be simplified as

$$\frac{d\overrightarrow{V_{out}}}{dt} + [\sigma(\beta+1) + j\Delta\omega]\overrightarrow{V_{out}} = -\frac{d\overrightarrow{V_{in}}}{dt} + [\sigma(\beta-1) + j\Delta\omega]\overrightarrow{V_{in}} \qquad (4)$$

The equation (3) can also be expressed by the transfer function as

$$\overrightarrow{V_{out}}(s) = \frac{-s + [\sigma(\beta-1) + j\Delta\omega]}{s + [\sigma(\beta+1) + j\Delta\omega]}\overrightarrow{V_{in}}(s) \qquad (5)$$

Due to the 3dB power divider between the input and output ports, the relationship between the input and output of the SLED for each cavity can be present respectively as

$$\vec{V}_{sled_out_cavity1}(s) = \frac{1}{\sqrt{2}} \frac{-s + [\sigma(\beta-1) + j\Delta\omega]}{s + [\sigma(\beta+1) + j\Delta\omega]} e^{j\frac{\pi}{2}} \frac{1}{\sqrt{2}} \vec{V}_{sled_in}(s) \qquad (6)$$

$$\vec{V}_{sled_out_cavity2}(s) = e^{j\frac{\pi}{2}} \frac{1}{\sqrt{2}} \frac{-s+[\sigma(\beta-1)+j\Delta\omega]}{s+[\sigma(\beta+1)+j\Delta\omega]} \frac{1}{\sqrt{2}} \vec{V}_{sled_in}(s) \qquad (7)$$

From the structure of the SLED as shown in Fig. 1, the output signal of the SLED is formed by the reflected signals of the two cavities. Based on the equations (6) and (7), the SLED model can be constructed as Fig.4

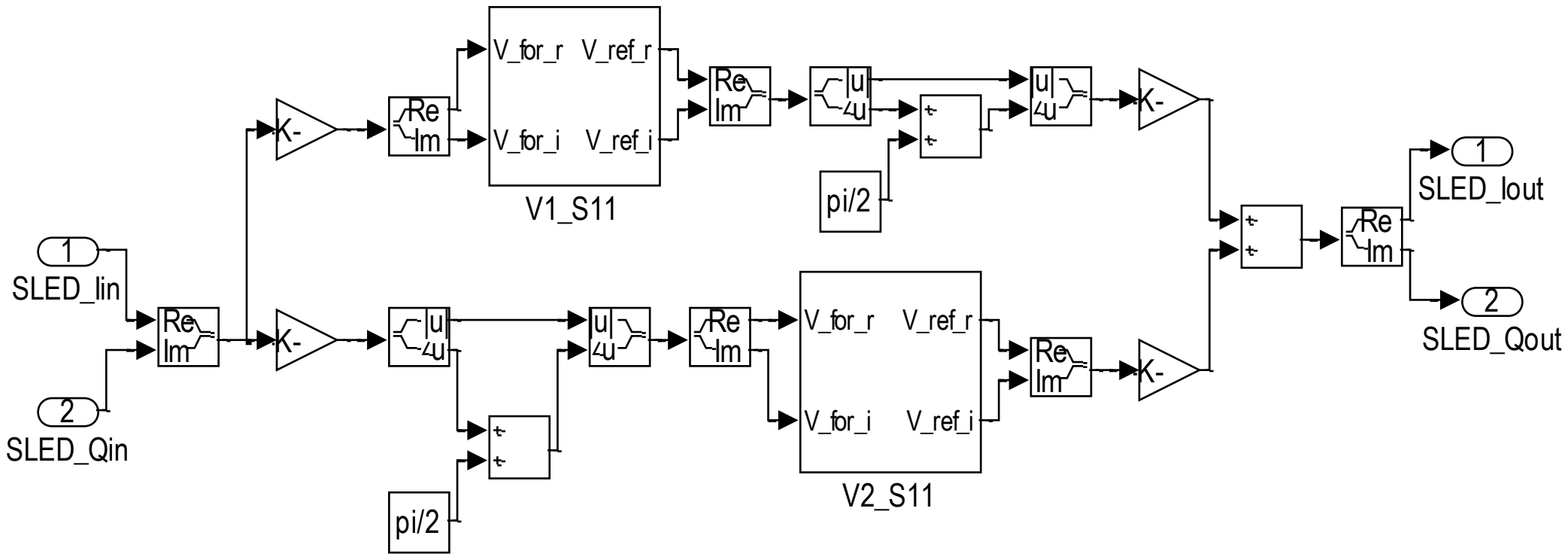

**Fig.4** The SLED mode

## 3 Results of simulation by SLED model

Based on previous theoretical analysis, we have carried out some simulations by the SLED model to optimize parameters.

### 3.1 Study of the working point of the SLED

The coupling coefficient $\beta$ and quality factor $Q_0$ of the cavity dominate the performance of the RF pulse compressor, such as the energy multiply factor and power efficiency. The parameters relevant to the calculation of the energy multiply factor and RF power efficiency is list in table 1. Using the parameters in table 1 and by tuning the coupling coefficient $\beta$ and quality factor $Q_0$, the tendency of the energy multiply factor and RF power efficiency can be mapped, as shown in Fig. 5. The power efficiency and the energy multiply increase with the quality factor and the input coupling coefficient.

**Table 1** Key parameters for Energy gain factor calculation

| | |
|---|---|
| RF frequency | 5712MHz |
| Accelerating structure Filling time | 372 ns |
| RF pulse length | 2.5 us |
| Reverse time | 2.0 us |

The optimal parameters are decided by the practical requirements. The point on the straight line in the Fig.5 shows the optimal operating point where both $\beta$ and $Q_0$ are small when the energy multiply factor reach the maximum value. According to the simulation results, the operating point is selected when quality factor $Q_0 = 160000$ and the coupling coefficient $\beta = 7.0$. And the corresponding energy multiply factor is 1.9029. Compared with the original working point ( $Q_0 = 180000$ , $\beta = 8.5$ ), the fabrication is easier and the energy multiplication factor was reduced by one 0.1%.

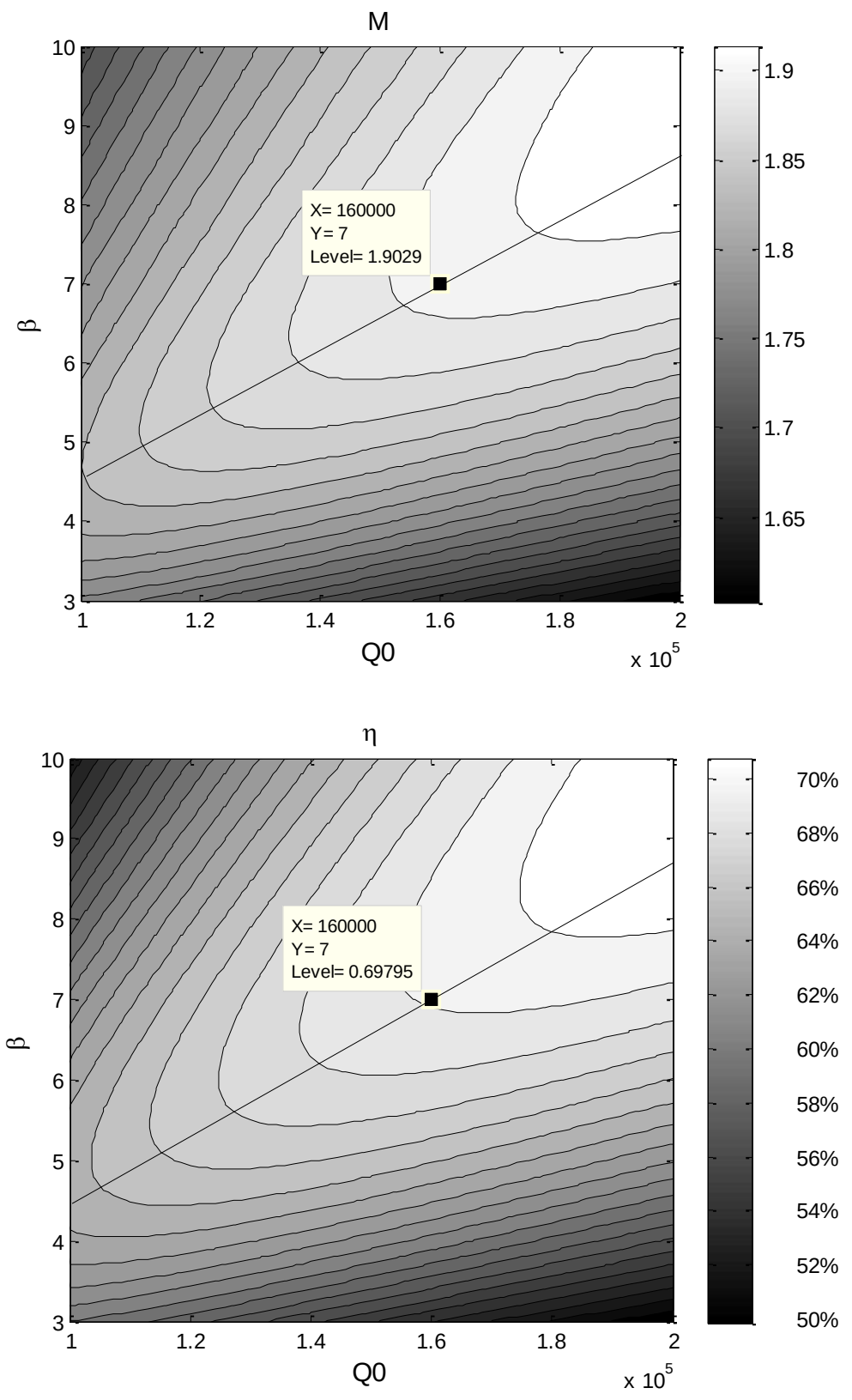

**Fig.5** The energy multiply factor and RF power efficiency mapped with the unloaded quality factor $Q_0$ and coupling coefficient

### 3.2 Course tuning before operation

In practice, more attentions are paid on the energy multiply factor. There are some frequency deviation from the operating frequency due to the temperature drift, some coupling coefficient deviation from the desired value and the unloaded quality factor error of the cavity due to machining tolerance. Using the SLED model, we can get the Energy multiplication factor change caused by the cavity frequency deviation, the input coupling coefficient deviation and the unloaded quality factor fluctuation, as shown in Fig.6. Fig. 6(a) and (b) shows that the energy multiplication factor declines with the increasing of frequency deviation and the coupling coefficient far from the desired value. The affect caused by the unloaded quality factor error of the cavity is shown in Fig. 6(c). The design value is not optimal values for getting the maximum energy multiplication factor which agrees well with analysis in the above section.

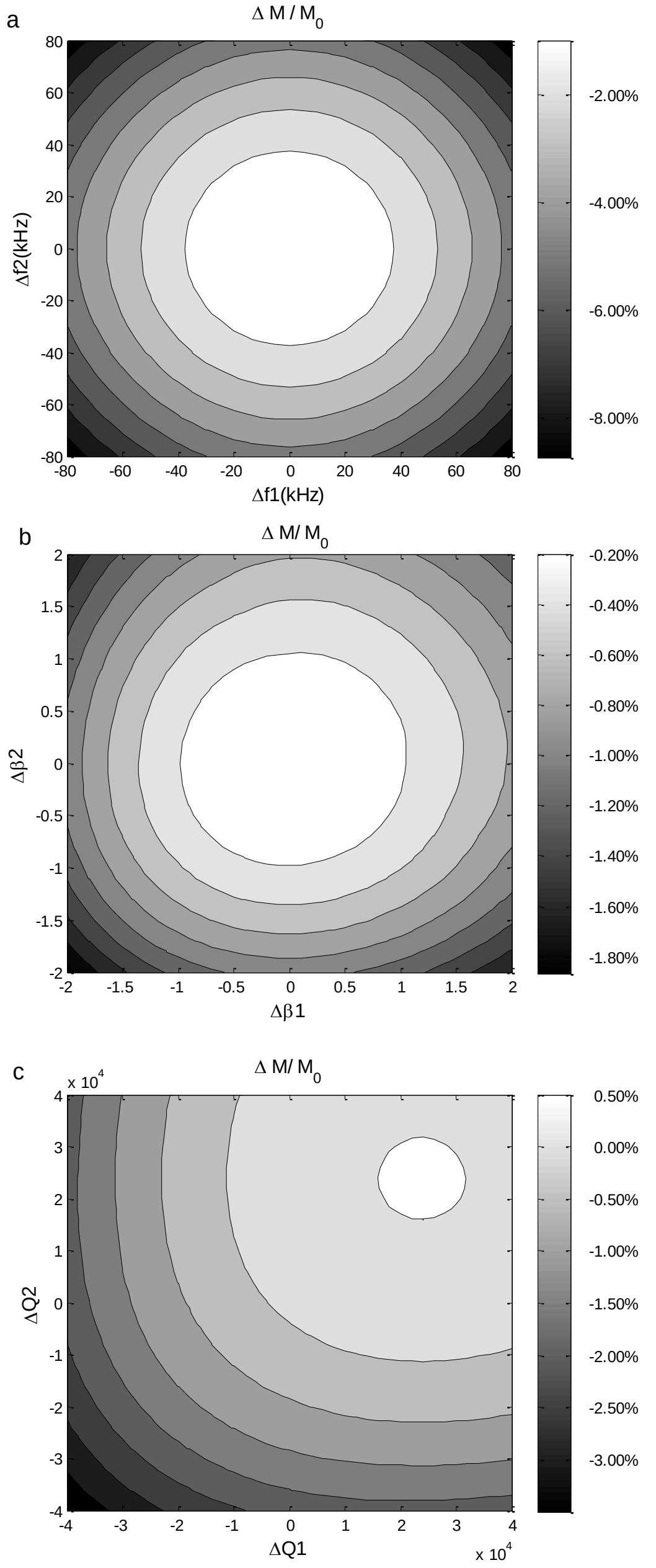

**Fig. 6** The Energy multiplication factor change with the cavity frequency deviation (a) , the input coupling coefficient deviation (b) and unloaded quality factor fluctuation ( $\beta = 7.0$ , $Q_0 = 160000$ ).

Before the actual operation, the RF pulse compressor should be tuned to maintained the energy multiply factor fluctuation less than 1%. According to the Fig.6, the frequency should be controlled in the range of $\pm 30$ kHz, the accuracy of the $\beta$ in the range of $\pm 2$, which is identical with the result in Fig. 5(a) and the $Q_0$ in the range of $\pm 2\mathrm{x}10^4$. Since the coupling coefficient $\beta$ and $Q_0$ cannot be tuned during the operation, so during the cold test they should be tuning in an optimal range in order to get a high precision, such as $\pm$ 0.25 for the coupling constant and $\pm 2\mathrm{x}10^3$ for $Q_0$ in order to attain the flatness of the energy gain factor 0.01%.

### 3.3 Fine tuning during operation

After cold test, the coupling coefficient and unloaded quality factor are tuned at the proper values and during the operation of the RF pulse compressor, they are fixed and cannot be tuned. There is only one parameter, frequency deviation, can be changed and tuned by controlling the temperature cooling water. During the operation, as the requirement of the flatness of the energy gain factor is less than 0.01%, the frequency of the cavity should be controlled within $\pm 0.2$ kHz. As the temperature expansion coefficient of the SLED cavity is about 106 kHz/$^0$C, the water temperature should be controlled within ±0.02 $^0$C.

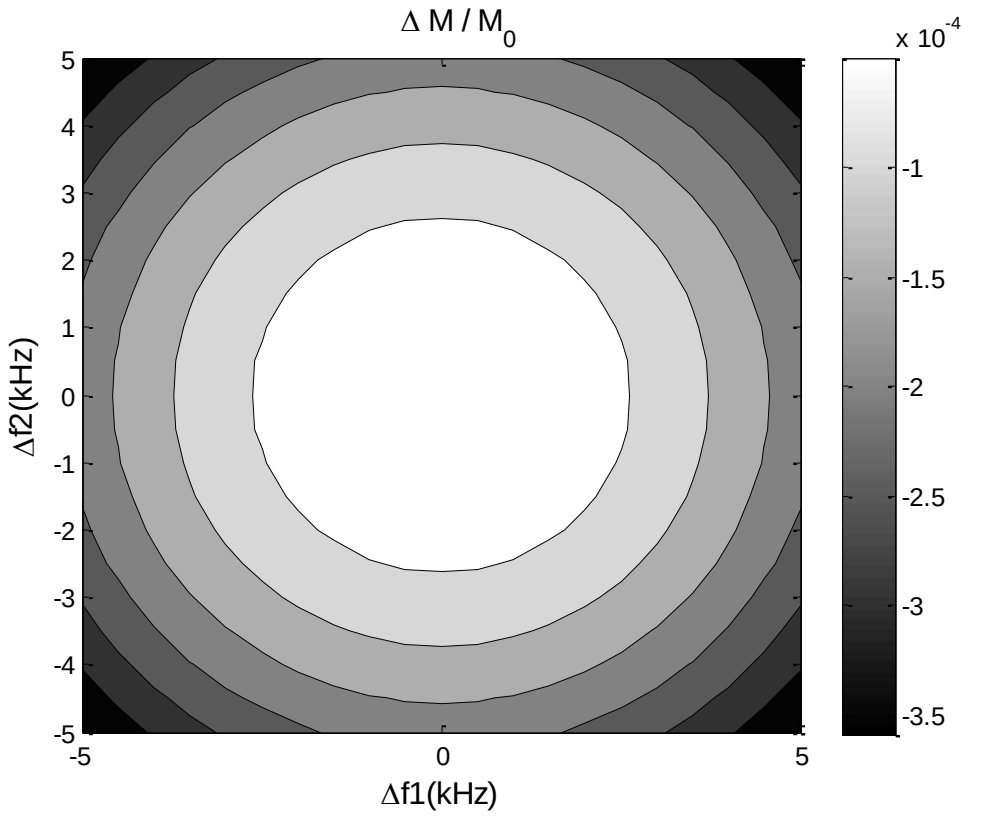

**Fig. 7** The Energy multiplication factor change with the cavity frequency deviation

According to the analysis above, the parameters requirements for getting energy multiplication factor flatness 0.01% are list in table 2.

**Table 2** parameter control range

| Parameters | Coarse tuning | Fine tuning |
|---|---|---|
| frequency deviation | ±30kHz | |
| temperature | | ±0.02 $^0$C (±2kHz) |
| coupling efficient deviation | ±0.25 | |
| unloaded quality factor difference | ±2x10$^4$ | |

## 4 Conclusion

RF compressor as a key technology for particle accelerators has been widely studied in many accelerator laboratories, such as KEK, CERN, IHEP and SLAC. But there are only a few researchers using equivalent circuit model to study the behavior of SLED and analyze the parameters. A detail process for building a mathematic model is shown in this paper. The simulation results and the analysis of the parameters deviation of the cavities are also presented. During our modeling analysis, the flatness of the energy gain factor 0.01% can be achieved with our model when the temperature of the cooling water $|T_c| \le 0.02^0 C$, then the maximum frequency detuning will be controlled in 2 kHz.

The SLED model can be used to make further study of the RF pulse compressor, more specific measurements will be carried out in the future.